\documentclass[11pt]{article}

\usepackage[french,english]{babel}
\usepackage[T1]{fontenc}

\usepackage{amsfonts,amsthm}
\usepackage{amsmath,mathtools}   
\usepackage{latexsym}
\usepackage{amssymb}
\usepackage{mathrsfs}
\usepackage{pifont}
\usepackage{tikz,ulem}
\usepackage[all]{xy}
\usepackage{hyperref}

\newtheoremstyle{newrem}{3pt}{3pt}{}{}
{\bfseries}{.}{.5em}{}

\newtheorem{theo}{Theorem}[section]

\newtheorem*{theo*}{Theorem}

\newtheorem{prop}[theo]{Proposition}

\theoremstyle{newrem}

\theoremstyle{definition}

\newtheorem*{term*}{Notation/Terminology}

\newcommand{\cD}{\mathcal{D}}

          \def\fp{{\mathfrak p}}

\newcommand{\ZZ}{{\mathbb Z}}

\newcommand{\pFq}[5]{{
{}_{#1}F_{#2}\left( \genfrac{}{}{0pt}{0}{#3}{#4}\middle| #5\right)
}}
\newcommand{\eps}{\varepsilon}

\newcommand{\arxiv}[1]{{\href{https://arxiv.org/abs/#1}{\texttt{arXiv:#1}}}}

\usepackage{geometry}
\newcommand{\doi}[1]{\href{https://doi.org/#1}{doi:#1}}
\numberwithin{equation}{section}

\begin{document}
\thispagestyle{empty}
\title{\bf Griffiths polynomials of Racah type
}
\author{
Nicolas Cramp\'e\textsuperscript{$1,2$},
Luc Frappat\textsuperscript{$2$},
Julien Gaboriaud\textsuperscript{$3$},
Eric Ragoucy\textsuperscript{$2$},
Luc Vinet\textsuperscript{$4,5$},
Meri Zaimi\textsuperscript{$4$}
\\[.9em]
\textsuperscript{$1$}
\small Institut Denis-Poisson CNRS/UMR 7013 - Universit\'e de Tours - Universit\'e
d'Orl\'eans,\\
\small~Parc de Grandmont, 37200 Tours, France.\\[.5em]
\textsuperscript{$2$}
\small Laboratoire d'Annecy-le-Vieux de Physique Th\'eorique LAPTh,\\
\small~Universit\'e Savoie Mont Blanc, CNRS, F-74000 Annecy,
 France.\\[.9em]
 \textsuperscript{$3$}
\small Graduate School of Informatics, Kyoto University, Sakyo-ku, Kyoto, 606-8501, Japan.\\[.9em]
 \textsuperscript{$4$}
\small Centre de Recherches Math\'ematiques, Universit\'e de Montr\'eal, P.O. Box 6128, \\
\small Centre-ville Station, Montr\'eal (Qu\'ebec), H3C 3J7, Canada.\\[.9em]
\textsuperscript{$5$}
\small IVADO, Montr\'eal (Qu\'ebec), H2S 3H1, Canada.\\[.9em]
}
\date{}
\maketitle

\bigskip\bigskip 

\begin{center}
\begin{minipage}{13cm}
\begin{center}
{\bf Abstract}\\
\end{center}
Bivariate Griffiths polynomials of Racah type are constructed from
univariate Racah polynomials. The bispectral properties of the former are deduced from simple properties of the latter. 
A duality relation and the orthogonality of these polynomials are provided. The domain of validity 
for the indices and variables of these polynomials is also determined. Particular limits on the parameters entering the polynomials allow to define several Griffiths polynomials of other types.
One special limit connects them to the original Griffiths polynomials (of Krawtchouk type).
Finally, a connection with the $9j$ symbols is made.
\end{minipage}
\end{center}

\medskip

\begin{center}
\begin{minipage}{14cm}

\textbf{MSC2020 database:} 33C80; 33C45; 16G60
\end{minipage}
\end{center}

\vfill

\textbf{E-mail:} crampe1977@gmail.com, luc.frappat@lapth.cnrs.fr, julien.gaboriaud@umontreal.ca, \\
eric.ragoucy@lapth.cnrs.fr, luc.vinet@umontreal.ca, meri.zaimi@umontreal.ca

\clearpage
\newpage

\section{Introduction}
This paper is dedicated to the characterization of the bivariate Griffiths polynomials of Racah type that have arisen in the construction performed in \cite{icosi}, of the representations of the rank 2 Racah algebra. The orthogonality, duality and bispectral properties are derived directly from the explicit defining expressions. These polynomials are two-variable generalizations of the univariate Racah polynomials \cite{Wilson, Koek} and have multivariate extensions. 
Although kindred to the bivariate (and multivariate) version of the Racah polynomials introduced by Tratnik \cite{Trat}, the Griffiths polynomials differ from the latter in that they are defined as convolutions of three instead of two univariate polynomials.

Our nomenclature is rooted in the story of the multivariate Krawtchouk polynomials. The univariate Krawtchouk polynomials \cite{Kra} are orthogonal with respect to the binomial distribution. In 1971, Griffiths \cite{Griff} introduced a multivariate version of these polynomials that are orthogonal, naturally, against the multinomial distribution (see also \cite{DC}). Their explicit expression in terms of Aomoto-Gel'fand series was subsequently provided in \cite{MT} out of considerations pertaining to association schemes and without reference to Griffiths. In 1991, as mentioned already, Tratnik introduced his multivariable analogs of the discrete families of the Askey tableau which include the Krawtchouk one \cite{Koek}. The bispectral properties of all these Tratnik polynomials were established in \cite{GI}. With Griffiths' multivariate Krawtchouk polynomials initially unbeknownst to them, Hoare and Rahman rediscovered these functions in a probabilistic context \cite{HR} roughly 15 years ago. This generated a lot of interest and studies by Rahman and others to the extent that for a certain period of time, these mathematical entities were referred to as Rahman polynomials. In time their paternity was reattributed to Griffiths however, the connection, if any, between the Griffiths and Tratnik multivariate Krawtchouk polynomials remained obscure for a while. This was resolved through the algebraic interpretation given in \cite{GVZ} showing that the bivariate Krawtchouk polynomials of Griffiths arise as matrix elements of the rotation group $SO(3)$ representation on energy eigenspaces of the three-dimensional isotropic harmonic oscillator (see also \cite{IT} and \cite{I12} for algebraic interpretations). Knowing that matrices representing rotations about a single axis have univariate Krawtchouk polynomials as entries \cite{NSU,Koorn} and remembering that in terms of Euler angles, a general 
three-dimensional
rotation is given as a product of three rotations about single 
axes, 
one understands that the bivariate Griffiths polynomials will be expressed as a convolution of three univariate Krawtchouk polynomials. With this understanding it is easy to see that the Tratnik polynomials correspond, in the Krawtchouk case, to setting one Euler angle to zero. This yields bivariate polynomials that are the product of only two entangled univariate polynomials and that are orthogonal with respect to the same trinomial distribution. One thus arrives at the notion of Griffiths and Tratnik polynomials of any given type by extending this feature regarding their composition in terms of three or two univariate polynomials respectively, to families of polynomials other than the Krawtchouk one. This carries beyond the bivariate situation in natural ways. Interestingly, already in 1997, Zhedanov had also found on his own these bivariate Krawtchouk polynomials of Griffiths as $9j$ symbols of the oscillator algebra \cite{Zhe}. The connection between this recoupling and the matrix element points of view is established in \cite{CVV}.

The Griffiths polynomials are of a recoupling coefficient nature. Arising typically as overlaps between representation bases, for instance in the tree method \cite{S, LVdJ}, within a realization they can be found as connection coefficients for families of polynomials. The Griffiths polynomials of concern here, namely those of Racah type, are directly connected to $9j$ coefficients as will be recalled below. They have indeed been studied and characterized in \cite{GV9j,S15,I17} as $9j$ symbols of $\mathfrak{su}(1,1)$ in the context of the generic superintegrable model on the $3$-sphere. The Griffiths polynomials of Racah type also appear as connection coefficients between two families of multivariate Jacobi polynomials related by a particular element of the permutation group \cite{IX}. In this latter paper, the orthogonality and the duality relations of these Griffiths polynomials have been already established.  
With these remarks we stress the differences between the Griffiths polynomials of Racah type and the multivariable Racah polynomials obtained by Stokman and van Diejen  \cite{SvD} from a truncation and $q \rightarrow 1$ limit of the symmetric Koornwinder-Macdonald polynomials based on root systems.

Overall, the reader will find in this text, derivations of all the basic properties of the Griffiths polynomials of Racah type and of their Tratnik counterparts as well. Starting from the explicit expressions, their fundamental features including orthogonality and bispectrality, will be straightforwardly obtained from the corresponding properties of the univariate constituents following techniques put forward in \cite{L-griff}. Novel contiguity relations will prove to be key ingredients. The paper will thus proceed as follows. Properties of the univariate polynomials are collected in the next section. Most are classical \cite{Koek} except for the contiguity relations that were recently obtained in \cite{icosi}. The reader should pay attention to the fact that for convenience a non-standard choice of normalization is made. Section \ref{sec:tratnik} is dedicated to the Tranik polynomials of Racah type. Four propositions that are proved in the spirit indicated above cover in turn the orthogonality, duality, recurrence relations and difference equations of these polynomials. Section \ref{sec:griffiths} is dedicated to the Griffiths polynomials of Racah type. Propositions bearing on these functions and similar to those of Section \ref{sec:tratnik} are provided and proved. It should be noted that the Griffiths and Tratnik polynomials have the same number of parameters. Relevant remarks relative to the polynomial character of the defining expression will be found. 
Attention is also paid to domains that might need to be further constrained when parameters are taken to be negative integers; the validity of the various properties are also confirmed in those instances. Finally, connections with the $9j$ symbols are discussed. In Section \ref{sect:limit}, various limits are seen to give other Griffiths polynomials, many of hybrid type; when all the parameters tend to infinity the multivariate Krawtchouk polynomials of Griffiths are recovered. The paper concludes with further questions of interest that this study raises.

\section{Properties of the univariate Racah polynomial}\label{sec:Rac}

We first recall the basic properties of the one-variable Racah polynomials.
These will be the tools that we will use to prove
the orthogonality, duality and bispectrality of the bivariate Racah
polynomials in the upcoming sections.

The Racah polynomials \cite{Wilson} 
are defined in terms of the generalized hypergeometric function as follows, for $N$ a non-negative integer and $n=0,1,\dots, N$,
\begin{align}\label{eq:2.1}
 R_n(\lambda(x);\alpha,\beta,\gamma,\delta)=
 \pFq{4}{3}{-n,\;n+\alpha+\beta+1, \;-x,\;x+\gamma+\delta+1}{\alpha+1,\; \beta+\delta+1,\;\gamma+1}{1}\,,
\end{align}
where $\lambda(x)=x(x+\gamma+\delta+1)$ and $\alpha$, $\beta+\delta$ or 
$\gamma$ is equal to $-N-1$. 
The functions \eqref{eq:2.1} are polynomials of degree $n$ 
with respect to the variable $\lambda(x)$.
In order to simplify the presentation, we shall use a normalization and a notation for the 
parameters that differ from the standard references:
\begin{align}\label{eq:p4F3}
\begin{aligned}
  p_n(x;c_1,c_2,c_3;N)=&\ \Omega(n;c_1,c_2,c_3;N) \, R_n(x(x+c_1+c_2+1);c_2,c_3,-N-1,N+1+c_1+c_2)\\
  =&\ \Omega(n;c_1,c_2,c_3;N) \,
\pFq{4}{3}{-n,\;n+c_2+c_3+1, \;-x,\;x+c_1+c_2+1}{c_2+1,\; N+2+c_1+c_2+c_3,\;-N}{1}\,,
\end{aligned}
\end{align}
where 
\begin{align}\label{def:Omega}
  \Omega(n;c_1,c_2,c_3;N)=&\binom{N}{n}
(2n+c_2+c_3+1)\frac{(c_2+1)_n(N+2+c_1+c_2+c_3)_n(c_1+1)_{N-n} }{ (c_3+1)_n (c_2+c_3+n+1)_{N+1}}\,. 
\end{align}
We used the Pochhammer symbols $(a)_n=a(a+1)\cdots(a+n-1)$.
To lighten the notation when there is no ambiguity, we shall write $p_n(x;N)\equiv p_n(x;c_1,c_2,c_3;N)$.
We will also use the notation $c_{ij}=c_i+c_j$ and $c_{ijk}=c_i+c_j+c_k$, 
for $i,j,k=0,1,\dots,4$.

\subsection{Duality, orthogonality and bispectral properties}

It is well-known that the Racah polynomials are solutions of a bispectral 
problem \textit{i.e.} that they verify a recurrence relation and a difference equation 
\cite{Wilson, Koek}. They are also self-dual and orthogonal. 
Let us recall these relations with the normalization used in this paper.

\paragraph{Duality.}  
Using invariance of the hypergeometric functions,
it is easily seen that
the Racah polynomials satisfy the following duality relation, 
for $n,x=0,1,\dots, N$:
\begin{align}\label{eq:dual}
 \Omega(x;c_3,c_2,c_1;N) \,p_n(x;c_1,c_2,c_3;N)=\Omega(n;c_1,c_2,c_3;N)\, p_x(n;c_3,c_2,c_1;N)\,.
\end{align}

\paragraph{Orthogonality.}  
The orthogonality relation with the normalization chosen in this paper reads, for $0\leq n,m\leq N$:
\begin{align}\label{eq:ortho}
\sum_{x=0}^N  \Omega(x;c_3,c_2,c_1;N) \,p_n(x;N)\,p_m(x;N)=\delta_{n,m}
\,\Omega(n;c_1,c_2,c_3;N)\,.
\end{align}

\paragraph{Recurrence relation.}
The Racah polynomials satisfy the following three-term
recurrence relation, for $n,x=0,1,\dots, N$:
\begin{subequations}\label{eq:rec-rac}
\begin{align}
\lambda(x)\, p_n(x;N)&=C_{n+1}(N)\, p_{n+1}(x;N)-\Sigma_n(N)\, p_n(x;N)+A_{n-1}(N)\,p_{n-1}(x;N)\,,
\end{align}
where $p_{-1}(x;N)=p_{N+1}(x;N)=0$ by convention and 
\begin{align}
&\lambda(x)\equiv\lambda(x;c_{12})=x(x+c_{12}+1)\,,\label{eq:lambda}\\
&A_n(N)\equiv A_n(c_1,c_2,c_3;N)=\frac{(n-N)(n+c_{123}+N+2)(n+c_2+1)(n+c_{23}+1)}{(2n+c_{23}+1)(2n+c_{23}+2)}\,,\label{eq:rec-racA}\\
&C_n(N)\equiv C_n(c_1,c_2,c_3;N)=\frac{n(n-c_1-N-1)(n+c_{23}+N+1)(n+c_3)}{(2n+c_{23})(2n+c_{23}+1)}\,,\label{eq:rec-racC}\\
&\Sigma_n(N)=A_n(N)+C_n(N)\,.\label{eq:a0}
\end{align}
\end{subequations}

\paragraph{Difference equation.}
The Racah polynomials obey the following second order difference equation, 
for $n,x=0,1,\dots, N$:
\begin{subequations}\label{eq:diff-rac}
 \begin{align}
 \mu_n\, p_n(x;N) &=B(x;N)\, p_{n}(x+1;N)-S(x;N)\, p_n(x;N)+D(x;N)\, p_n(x-1;N)\,,
\end{align}
where
\begin{align}
& \mu_n\equiv\mu_n(c_{23})=n(n+c_{23}+1)\,,\label{eq:mu0}\\
 &B(x;N)\equiv B(x;c_1,c_2,c_3;N)=\frac{(x-N)(x+c_2+1)(x+c_{123}+N+2)(x+c_{12}+1)}{(2x+c_{12}+1)(2x+c_{12}+2)}\,,\label{eq:diff-racB}\\
 &D(x;N)\equiv D(x;c_1,c_2,c_3;N)=\frac{x(x+c_1)(x-c_3-N-1)(x+c_{12}+N+1)}{(2x+c_{12})(2x+c_{12}+1)}\,\label{eq:diff-racD},
 \\
 &S(x;N)=B(x;N)+D(x;N)\,.\label{eq:diff-racS}
\end{align}
\end{subequations}
Note that, due to our choice of normalization for the Racah 
polynomials, the position of the coefficients $A_n$ and $C_n$ in the 
recurrence relation is inverted w.r.t.~\cite{Koek}. 
Since our normalization does not depend on the variable $x$, the 
difference equation is not affected.

\subsection{Contiguity relations}

In the upcoming proofs, we will make use of relations relating Racah 
polynomials of a given $N$ to the ones with shifted parameter $N\pm1$. 
These are called contiguity relations 
and have been proven in the Appendix of \cite{icosi}.
We recall them below in the normalization used in this paper.
For convenience, we first introduce the function $F(x;c_1,c_2)$:
\begin{equation}\label{eq:defF}
F(x;c_1,c_2)=\frac{(x+c_2+1)(x+c_{12}+1)}{(2x+c_{12}+1)(2x+c_{12}+2)}\,.
\end{equation}

\paragraph{Contiguity recurrence relations.}
The first relations relate various polynomials with shifted degree $n$. 
By analogy with the three-term recurrence relation, we call these 
the contiguity recurrence relations. 
These read as follows:
\begin{subequations}\label{eq:contiguity_recurrence}
\begin{align}\label{eq:cont-rec-rac}
\lambda^\pm(x;N) p_n(x;N)&=C^\pm_{n+1}(N)\, p_{n+1}(x;N\pm 1)
-\Sigma^\pm_n(N)\, p_n(x;N\pm1)+A^\pm_{n-1}(N)\,p_{n-1}(x;N\pm 1)\,,
\end{align}
where 
\begin{align}
&\lambda^+(x;N)\equiv \lambda^+(x;c_{12};N)=(x+c_{12}+N+2)(x-N-1)\,,\\
&A^+_{n}(N)\equiv A^+_n(c_2,c_3;N)=-(n-N-1)(n-N)\,F(n;c_3,c_2) \, \\
& C^+_{n}(N)\equiv C^+_{n}(c_2,c_3;N)=A^+_{-n-c_{23}-1}(N)\,,\\
&\Sigma^+_n(N)=A^+_n(N)+C^+_n(N)+(N+1)(N+1+c_3)\,,\label{eq:2.9e}\\[1ex]
&\lambda^-(x;N)\equiv \lambda^-(x;c_{123},c_3;N)=(x+c_{123}+N+1)(x-N-c_3)\,,\\
& A^-_n(N)\equiv A^-_n(c_1,c_2,c_3;N)=-(n+c_{123}+N+1)(n+c_{123}+N+2)\,F(n;c_3,c_2)\,, \\
&  C^-_n(N)\equiv C^-_n(c_1,c_2,c_3;N)=A^-_{-n-c_{23}-1}(N)\,,\\
&\Sigma^-_n(N)=A^-_n(N)+C^-_n(N)+(N+c_{12}+1)(N+c_{123}+1)\,.
\end{align}
\end{subequations}

\paragraph{Contiguity difference equations.}
The second set of relations relate various polynomials 
with shifted variable $x$. 
By analogy with the difference equation, we call these 
the contiguity difference equations. 
These read as follows:
\begin{subequations}
\begin{align}\label{eq:cont-diff-rac}
\mu_n^\pm(N)\, p_n(x;N) &=B^\pm(x;N)\, p_{n}(x+1;N\pm 1)
-S^\pm(x;N)\, p_n(x;N\pm 1)+D^\pm(x;N) p_n(x-1;N\pm 1)\,,
\end{align}
where
\begin{align}
&\mu^+_n(N)\equiv \mu^+_n(c_{123},c_1;N)=(n+c_{123}+N+2)(n-N-1-c_1)\,,\\
&B^+(x;N)\equiv B^+(x;c_1,c_2,c_3;N)= -(x+c_{123}+N+2)(x+c_{123}+N+3)\,F(x;c_1,c_2)\,,\\
& D^+(x;N)\equiv D^+(x;c_1,c_2,c_3;N)=
B^+(-x-c_{12}-1;N)\,,\\
&S^+(x;N)=B^+(x;N)+D^+(x;N)+(c_{23}+N+2)(c_{123}+N+2)\,,\\[1ex]
&\mu^-_n(N)\equiv \mu^-_n(c_{23};N)=(n-N)(n+c_{23}+N+1)\,,\\
&B^-(x;N)\equiv B^-(x;c_1,c_2;N)=-(x-N)(x-N+1)\,F(x;c_1,c_2)\,, \\
& D^-(x;N)\equiv D^-(x;c_1,c_2;N)=
B^-(-x-c_{12}-1;N)\,,\\
&S^-(x;N)=B^-(x;N)+D^-(x;N)+N(c_1+N)\,.
\end{align}
\end{subequations}

\section{Tratnik polynomials of Racah type}\label{sec:tratnik}
A generalization of the Racah polynomials  has been introduced in \cite{Trat} and
will be called the Tratnik polynomials of Racah type.
These can be written as a product of two univariate Racah
polynomials:
\begin{align}\label{eq:tratnik}
 T_{i,j}(x,y)=p_i(x;c_1,c_2,c_3;N-j)\,p_j(y;c_3,c_0,c_4;N-x)\,,
\end{align}
for $i,j,x,y\in \ZZ_{\geq 0}$ such that $i+j,x+y\leq N$ and
where the parameters obey the constraint:
\begin{align}\label{eq:contrainte}
 c_0+c_1+c_2+c_3+c_4=-2N-3\,.
\end{align}
The notation $T_{i,j}(x,y)$ is short for $T_{i,j}(x,y;c_1,c_2,c_3,c_4;N)$.
From the convention for the Racah polynomials, one gets: 
$T_{-1,j}(x,y)=T_{i,-1}(x,y)=T_{i,N+1-i}(x,y)=0$.
To keep it simple, when there are no ambiguities we will call these 
$T_{i,j}(x,y)$ Tratnik polynomials.

Let us emphasize that we can multiply the previous expression of 
$T_{i,j}(x,y)$ by any function of type $f(i,j)$ or $g(x,y)$. 
Doing so, the relations satisfied by $T_{i,j}(x,y)$ will still hold but 
with modified explicit expressions of their coefficients.

\subsection{Polynomiality}
Definition \eqref{eq:tratnik} of $T_{i,j}(x,y)$ is only valid 
when $x$, $y$ are non-negative integers and $x+y\leq N$.
This triangular domain, called the grid, is the domain over which
the functions are orthogonal with respect to each other 
(see \eqref{eq:orthoT} below).
%
It is possible to multiply $T_{i,j}(x,y)$ by some appropriate factors 
$f(i,j)$ and $g(x,y)$ so that
it becomes a function well-defined not only on the grid
but also for $x$, $y$ real, complex numbers or even formal variables.
Below we show precisely that, 
up to transformations only valid on the grid,
this function becomes a polynomial.

Starting from definition \eqref{eq:tratnik} and using properties of the Pochhammer symbols, $T_{i,j}(x,y)$ can be rewritten as follows:
\begin{align}\label{eq:tratnik2}
    T_{i,j}(x,y)&=\Omega(i;c_1,c_2,c_3;N-j)\,\frac{(2j+c_{04}+1)(c_0+1)_j(c_{04}+1)_j}{(c_4+1)_j}\, \frac{(c_3+1)_{N-x}}{(c_{04}+1)_{N-x+1}}\\
    &\times\frac{(x-N)_j(-c_{12}-N-x-1)_j}{(x-N-c_3)_j(-c_{123}-N-x-1)_j}
    \pFq{4}{3}{-i,\; i+c_{23}+1,\; -x,\; x+c_{12}+1 }
          {c_2+1,\; c_{123}+N+2-j,\; j-N}{1}\nonumber\\
       &\times   \pFq{4}{3}{-j,\; j+c_{04}+1,\; -y,\; y+c_{03}+1 }
          {c_0+1,\; -c_{12}-N-x-1,\; x-N}{1}  \,.\nonumber
\end{align}
The factors in the first line of the above expression depend only on $(i,j)$ or on $x$. As explained previously, $T_{i,j}(x,y)$ divided by these factors  satisfies the same relations as $T_{i,j}(x,y)$ but 
with modified coefficients; therefore, we can focus on the last two lines of \eqref{eq:tratnik2}. These define a rational function w.r.t.~$\lambda(x;c_{12})$ and $\lambda(y;c_{03})$.
More precisely, the denominator is a polynomial of degree $j$ w.r.t.~$\lambda(x;c_{12})$, while the numerator is a product of a polynomial of degree $i$ w.r.t.~$\lambda(x;c_{12})$ and a bivariate polynomial of total degree $j$ w.r.t.~$\lambda(x;c_{12})$ and $\lambda(y;c_{03})$.

To rewrite this as a polynomial instead of a rational function, 
it is necessary to use two successive Whipple's transformations.
We recall that the Whipple's transformation 
(see \textit{e.g.}~relation (2.10.5) in \cite{GR}) reads as 
\begin{align}
      \pFq{4}{3}{-n,\;a, \;b,\;c }
          {d,\; e,\; f}{1}=
          \frac{(e-a)_n(f-a)_n}{(e)_n(f)_n}
          \pFq{4}{3}{-n,\;a, \;d-b,\;d-c }
          {d,\; 1+a-e-n,\; 1+a-f-n}{1}\,,
\end{align}
where $n$ is an integer and $a+b+c+1=d+e+f+n$. 
First apply Whipple's transformation on the hypergeometric function 
${}_4F_3$ in the second line of \eqref{eq:tratnik2} 
with the identifications $n\to x,\ a\to x+c_{12}+1,\ b\to -i,\ c\to i+c_{23}+1,\ d\to j-N,\ e\to c_2+1, \ f \to c_{123}+N+2-j$ to get
\begin{align}\label{eq:Wh1}
    \frac{(-x-c_1)_x(c_3+N-j-x+1)_x}{(c_2+1)_x(c_{123}+N-j+2)_x}\pFq{4}{3}{i+j-N,\;j-N-i-c_{23}-1,\;-x, \;x+c_{12}+1 }
          {c_1+1,\; j-c_3-N,\; j-N}{1}\,.
\end{align}
Let us emphasize that this transformation is only possible 
when $x$ is on the grid.
The ${}_4F_3$ in \eqref{eq:Wh1} is then further modified using 
Whipple's transformation with the identifications $n\to N-i-j,\ a\to j-N-i-c_{23}-1,\ b\to -x,\ c\to x+c_{12}+1,\ d\to j-N,\ e\to c_1+1,\ f\to j-c_3-N$ to get
\begin{align}
    &\frac{(c_{123}+N+2+i-j)_{N-i-j} (i+c_2+1)_{N-i-j} }{(j-c_3-N)_{N-i-j} (c_1+1)_{N-i-j}}\nonumber\\
    \times&\pFq{4}{3}{i+j-N,\;j-N-i-c_{23}-1,\;x-N+j, \;-N-x+j-c_{12}-1 }
          {2j+c_{04}+2,\; j-c_2-N,\; j-N}{1}\,.
\end{align}
Using properties of the Pochhammer symbols as well as Whipple
transformations that are valid on the grid, 
$T_{i,j}(x,y)$ can thus be rewritten as 
\begin{align}\label{eq:Tratpol}
    T_{i,j}(x,y)&=\frac{(-1)^{i+j}}{j!}\binom{N-j}{i}\frac{(2i+c_{23}+1)(c_0+1)_j(c_2+1)_{N-j}(c_1+1)_x}
    {(c_{23}+i+1)_{N-j+1}(c_{04}+j+1)_j(c_4+1)_j(c_2+1)_x}\nonumber\\
   &\times        \pFq{4}{3}{i+j-N,\;-N-i+j-c_{23}-1,\;x-N+j, \;-N-x+j-c_{12}-1}           {2j+c_{04}+2,\; j-c_2-N,\; j-N}{1}
\nonumber\\
    &\times(x-N)_j(-c_{12}-N-x-1)_j\ 
          \pFq{4}{3}{-j,\;j+c_{04}+1,\;-y,\;y+c_{03}+1 }
          {c_0+1,\; -c_{12}-N-x-1,\; x-N}{1}\,.
\end{align}
With this expression of $T_{i,j}(x,y)$, one sees that 
the last two lines are a product of two polynomials: 
a bivariate polynomial of total degree $j$ w.r.t.~$\lambda(x;c_{12})$ 
and $\lambda(y;c_{03})$ multiplied by 
a polynomial of degree $N-i-j$ w.r.t.~$\lambda(x;c_{12})$.

\subsection{Connection with the historical notations}

In \cite{Trat}, Tratnik introduced generalizations of the Racah polynomials in $p$ variables which reads for $p=2$ as follows, 
with $0\leq x_1 \leq x_2\leq -\gamma-1$:
\begin{align}
&R\left(
\genfrac{}{}{0pt}{0}{\alpha_1,\;\alpha_2,\;\alpha_3}{\eta,\;\gamma}
\middle|
\genfrac{}{}{0pt}{0}{x_1,\;x_2}{n_1,\;n_2}
\right)
= (\eta+1)_{n_1}\,(\alpha_1+\alpha_2+x_2)_{n_1}\,(-x_2)_{n_1}\,
\pFq{4}{3}{-n_1,\;n_1+\alpha_2+\eta, \;-x_1,\;x_1+\alpha_1 }
          {\eta+1,\; \alpha_1+\alpha_2+x_2,\; -x_2}{1}\nonumber\\
&\qquad\times\,
(2n_1+\eta+\alpha_2+1)_{n_2}\,(n_1+\alpha_1+\alpha_2+\alpha_3-\gamma-1)_{n_2}\,(n_1+\gamma+1)_{n_2}\nonumber\\
&\qquad\times\,
\pFq{4}{3}{-n_2,\; n_2+2n_1+\eta+\alpha_2+\alpha_3, \; -x_2+n_1,\;
           x_2+n_1+\alpha_1+\alpha_2}
          {2n_1+\eta+\alpha_2+1,\; n_1+\alpha_1+\alpha_2+\alpha_3-
           \gamma-1,\;n_1+\gamma+1}{1}\,.\label{eq:historical}
\end{align}
In order to recover the previous expression \eqref{eq:Tratpol} for Tratnik polynomials, let us write
\begin{gather}
\begin{gathered}
    \gamma = -N-1\,,\qquad x_1 = y\,,\qquad x_2 = N-x\,,\qquad 
     n_1 = j\,,\qquad n_2 = N-i-j\,,\\
    \eta = c_0\,,\qquad \alpha_1 = c_0+c_3+1\,,\qquad 
     \alpha_2 = c_4+1\,,\qquad \alpha_3 = c_1+1\,,
\end{gathered}
\end{gather}
Finally, an explicit computation leads to the following relation
\begin{align}
R\left(
\genfrac{}{}{0pt}{0}{\alpha_1,\;\alpha_2,\;\alpha_3}{\eta,\;\gamma}
\middle|
\genfrac{}{}{0pt}{0}{x_1,\;x_2}{n_1,\;n_2}
\right)
&= 
 \frac{(-1)^{i+j}j!(N-j-i)!(c_4+1)_j(i+c_{23}+1)_{N-j+1} (j+c_{04}+1)_{N-i+1}}{(c_2+1)_i(2i+c_{23}+1) (2j+c_{04}+1)}
\nonumber\\
&\times\frac{(c_2+1)_x}{(c_1+1)_x}\, T_{i,j}(x,y)\,.
\end{align}

\subsection{Orthogonality and duality} 
We provide here the orthogonality relation satisfied by the 
Tratnik polynomials with our conventions. 
\begin{prop}[\textbf{Orthogonality}]
The Tratnik polynomials satisfy the orthogonality relation
\begin{align}\label{eq:orthoT}
\begin{aligned}
&\sum_{\genfrac{}{}{0pt}{}{0\leq x,y\leq N}{x+y\leq N}}
\Lambda(x;c_1,c_2;N)\,\Omega(y;c_4,c_0,c_3;N-x)\, 
T_{i,j}(x,y)\, T_{k,\ell}(x,y) \\
&\quad =\delta_{i,k}\,\delta_{j,\ell}\,\Lambda(j;c_4,c_0;N)\,\Omega(i;c_1,c_2,c_3;N-j)\,,
\end{aligned}
\end{align}
where the function $\Omega$ is given by \eqref{def:Omega} and 
\begin{equation}\label{def:Lambda}
\Lambda(x;c_1,c_2;N)=(-1)^x\binom{N}{x}(2x+c_{12}+1)\frac{(c_2+1)_x}{(c_1+1)_x(x+c_{12}+1)_{N+1}} \,.
\end{equation}
\end{prop}
\proof 
Using the relation $c_0+c_1+c_2+c_3+c_4=-2N-3$, one proves by direct 
computation that
\begin{equation}\label{eq:LO}
\frac{\Lambda(x;c_1,c_2;N)}{ \Lambda(j;c_4,c_0;N)}
=\frac{\Omega(x;c_3,c_2,c_1;N-j)}{\Omega(j;c_3,c_0,c_4;N-x)}\,.
\end{equation}
One then uses this property to rewrite \eqref{eq:orthoT} as 
\begin{align}
\sum_{x=0}^N \sum_{y=0}^{N-x}   
\frac{\Omega(x;c_3,c_2,c_1;N-j)\,\Omega(y;c_4,c_0,c_3;N-x)}        
     {\Omega(j;c_3,c_0,c_4;N-x)\,\Omega(i;c_1,c_2,c_3;N-j)}
\,T_{i,j}(x,y)\, T_{k,\ell}(x,y)=\delta_{i,k}\,\delta_{j,\ell}\,.
\end{align}
This last expression is easily checked by 
using the expression  \eqref{eq:tratnik} of  Tratnik polynomials 
in terms of the univariate polynomials  and using 
their orthogonality relation \eqref{eq:ortho}.
\endproof

As was the case for Racah polynomials, 
the duality of the Tratnik polynomials can be expressed in terms
of the normalization.
\begin{prop}[\textbf{Duality}]
The Tratnik polynomials obey the following duality relation for $i,j,x,y$ 
non-negative integers such that $i+j\leq N$ and $x+y\leq N$:
\begin{equation}\label{eq:dualT}
\frac{T_{i,j}(x,y;c_1,c_2,c_3,c_4;N)}
     {\Omega(i;c_1,c_2,c_3;N-j)\,\Lambda(j;c_4,c_0;N)}=
\frac{T_{y,x}(j,i;c_4,c_0,c_3,c_1;N)}
     {\Omega(y;c_4,c_0,c_3;N-x)\,\Lambda(x;c_1,c_2;N)}\,.
\end{equation}
\end{prop}
\proof 
This is a direct calculation by using the duality relation for 
the univariate Racah polynomials as well as equation \eqref{eq:LO}.
\endproof

\subsection{Bispectral properties}
The bispectrality of these bivariate polynomials was established in \cite{GI} and its relation with the representations of the Kohno--Drinfeld Lie algebra was explained in \cite{I17}.
This is summarized in the following propositions. 
To the best of our knowledge, the proofs provided below are new. 

\begin{prop}[\textbf{Recurrence relations}]
\label{prop:rec-trat}
The Tratnik polynomials, 
for $x,y,i,j$ non-negative integers and $x+y\leq N$, $i+j\leq N$,
obey the following recurrence relations:
\begin{align}\label{eq:rec1-Trat}
\begin{aligned}
\lambda(x;c_{12})\,T_{i,j}(x,y)&= C_{i+1}(c_1,c_2,c_3;N-j)\,T_{i+1,j}(x,y) 
+A_{i-1}(c_1,c_2,c_3;N-j)\,T_{i-1,j}(x,y)\\
&\quad-\Sigma_{i}(c_1,c_2,c_3;N-j)\,T_{i,j}(x,y)\,,
\end{aligned}
\end{align}
where the coefficients $A_i$, $C_i$, $\Sigma_i$ are given by 
\eqref{eq:rec-racA}, \eqref{eq:rec-racC}, \eqref{eq:a0},
and 
\begin{subequations}
\label{def:Aij}
\begin{equation}\label{eq:rec2-trat}
\Big(\lambda(y;c_{30})+\tfrac{1}{2}(c_3+1)(c_0+1)\Big)\,T_{i,j}(x,y) 
= \sum_{\eps,\eps'=0,\pm1}A^{\eps,\eps'}_{i+\eps,j+\eps'}\,
  T_{i+\eps,j+\eps'}(x,y)\,,
\end{equation}
with
\begin{align}
 A_{i,j}^{+,+} &= -F(-j-c_{04}-1;c_4,c_0)\,C^-_{i}(c_1,c_2,c_3;N-j+1)\,,
 \label{def:App}\\
 A_{i,j}^{0,+} &= \phantom{-}F(-j-c_{04}-1;c_4,c_0)\,\Sigma^-_{i}(c_1,c_2,c_3;N-j+1)\,,\label{def:A0p}\\
 A_{i,j}^{-,+} &=   -F(-j-c_{04}-1;c_4,c_0)\,A^-_{i}(c_1,c_2,c_3;N-j+1)\,,\label{def:Amp}
\\
 A_{i,j}^{+,-} &=   -F(j;c_4,c_0)\,C^+_{i}(c_2,c_3;N-j-1)\,,\\
A_{i,j}^{0,-} &= \phantom{-}F(j;c_4,c_0)\,\Sigma^+_{i}(c_2,c_3;N-j-1)\,,
\label{def:A0m}\\
 A_{i,j}^{-,-} &= - F(j;c_4,c_0)\,A^+_{i}(c_2,c_3;N-j-1)\,,\label{def:Amm}
 \\
A_{i,j}^{+,0} &= \left[F(j;c_4,c_0)+F(-j-c_{04}-1;c_4,c_0)\right]\,C_{i}(c_1,c_2,c_3;N-j)\,,\\
A_{i,j}^{0,0} &=-\left[F(j;c_4,c_0)+F(-j-c_{04}-1;c_4,c_0)\right]\,\\
&\qquad\times\left[ \Sigma_{i}(c_1,c_2,c_3;N-j)
+(N-j)^2+(c_{123}+2)(N-j)
+\tfrac{1}{2}(c_3+1)(c_{123}+1) \right]\,,\nonumber\\
A_{i,j}^{-,0} &= \left[F(j;c_4,c_0)+F(-j-c_{04}-1;c_4,c_0)\right]\,A_{i}(c_1,c_2,c_3;N-j)\,,\label{def:Am0}
\end{align}
\end{subequations}
and $F(x;c_1,c_2)$ defined in \eqref{eq:defF}.
\end{prop}
\proof
The equation \eqref{eq:rec1-Trat} is a direct consequence of the recurrence relation \eqref{eq:rec-rac} applied to the Racah polynomial $p_i(x;c_1,c_2,c_3;N-j)$ in $T_{i,j}(x,y)$.

To prove \eqref{def:Aij}, we apply the recurrence relation to the second polynomial $p_j(y;c_3,c_0,c_4;N-x)$ such that the l.h.s.~of \eqref{eq:rec2-trat} becomes
\begin{align}
\begin{aligned}
p_i(x;c_1,c_2,c_3;N-j)&\Big(
C_{j+1}(c_3,c_0,c_4;N-x)\,p_{j+1}(y;c_3,c_0,c_4;N-x)\Big.\\
&~-\left[\Sigma_{j}(c_3,c_0,c_4;N-x)-\tfrac{1}{2}(c_3+1)(c_0+1)\right]\,p_{j}(y;c_3,c_0,c_4;N-x)\\
&~\Big. +A_{j-1}(c_3,c_0,c_4;N-x)\,p_{j-1}(y;c_3,c_0,c_4;N-x)
\Big)\,.
\end{aligned}
\end{align}
Then, remarking that
\begin{align}
&C_{j+1}(c_3,c_0,c_4;N-x)= 
 -F(-j-c_{04}-2;c_4,c_0)\,\lambda^-(x;c_{123},c_3;N-j)\,,\\[1ex]
&\begin{aligned}
&\left[\Sigma_{j}(c_3,c_0,c_4;N-x)-\tfrac{1}{2}(c_3+1)(c_0+1)\right]
 =-\left[F(j;c_4,c_0)+F(-j-c_{04}-1;c_4,c_0)\right]\,\\
&\qquad \times\left[ \lambda(x;c_{12})-(N-j)^2-(c_{123}+2)(N-j)
 -\tfrac{1}{2}(c_3+1)(c_{123}+1)  \right]\,,
\end{aligned}\\[1ex]
&A_{j-1}(c_3,c_0,c_4;N-x)=  -F(j-1;c_4,c_0)\,\lambda^+(x;c_{12};N-j)\,,
\end{align}
we apply the recurrence relation \eqref{eq:rec-rac} and the contiguity recurrence relation \eqref{eq:cont-rec-rac} to the first polynomial $p_i(x;c_1,c_2,c_3;N-j)$. 
This allows us to recognize Tratnik polynomials with shifted indices, 
and we obtain the r.h.s.~of \eqref{eq:rec2-trat} with the coefficients 
given in \eqref{def:App}--\eqref{def:Am0}.
\endproof

\begin{prop}[\textbf{Difference equations}]
The Tratnik polynomials, 
for $x,y,i,j$ non-negative integers and $x+y\leq N$, $i+j\leq N$, satisfy the following two difference equations:
\begin{align}
\begin{aligned}
\mu_j(c_{04})\,T_{i,j}(x,y) &= B(y;c_3,c_0,c_4;N-x)\,T_{i,j}(x,y+1) 
+D(y;c_3,c_0,c_4;N-x)\,T_{i,j}(x,y-1)\\
&\quad-S(y;c_3,c_0,c_4;N-x)\,T_{i,j}(x,y) \,,
\end{aligned}
\end{align}
where the coefficients $B$, $D$, $S$ are given by 
\eqref{eq:diff-racB}, \eqref{eq:diff-racD}, \eqref{eq:diff-racS}, and
\begin{subequations}
\label{def:Dij}
\begin{equation}
\Big(\mu_i(c_{23})+\tfrac{1}{2}(c_2+1)(c_3+1)\Big)\,T_{i,j}(x,y) = \sum_{\eps,\eps'=0,\pm1}D^{\eps,\eps'}(x,y)\,T_{i,j}(x+\eps,y+\eps')\,,
\end{equation}
where 
\begin{align}
D^{+,+}(x,y) &= -F(x;c_1,c_2)\,B^-(y;c_3,c_0;N-x)\,,\\
D^{+,0}(x,y) &= \phantom{-}F(x;c_1,c_2)\,S^-(y;c_3,c_0;N-x)\,,\\
D^{+,-}(x,y) &= -F(x;c_1,c_2)\,D^-(y;c_3,c_0;N-x)\,, \\
D^{-,+}(x,y) &= -F(-x-c_{12}-1;c_1,c_2)\,B^+(y;c_3,c_0,c_4;N-x)  \,,\\
D^{-,0}(x,y) &= \phantom{-}F(-x-c_{12}-1;c_1,c_2)\,S^+(y;c_3,c_0,c_4;N-x) 
\,,\\
D^{-,-}(x,y) &= -F(-x-c_{12}-1;c_1,c_2)\,D^+(y;c_3,c_0,c_4;N-x)\,,\\
D^{0,+}(x,y) &= \left[F(x;c_1,c_2) +F(-x-c_{12}-1;c_1,c_2)\right]\,
 B(y;c_3,c_0,c_4;N-x)\,\,,\\
D^{0,0}(x,y) &= -\left[F(x;c_1,c_2) +F(-x-c_{12}-1;c_1,c_2)\right]\, \\
&\qquad\times\left[S(y;c_3,c_0,c_4;N-x) +(N-x)^2+(c_{034}+2)(N-x)
 +\tfrac{1}{2}(c_3+1)(c_{034}+1)\right]\,,\nonumber\\
D^{0,-}(x,y) &=\left[F(x;c_1,c_2) +F(-x-c_{12}-1;c_1,c_2)\right]\,
 D(y;c_3,c_0,c_4;N-x)\,.
\end{align}
\end{subequations}
\end{prop}
\proof
The calculation proceeds along the same lines as what was done 
for Proposition \ref{prop:rec-trat}.
\endproof

\section{Griffiths polynomials of Racah type}\label{sec:griffiths}

The Griffiths polynomials of Racah type are defined as follows, for $i,j,x,y$ 
non-negative integers such that $i+j,x+y\leq N$:
\begin{align}
 G_{i,j}(x,y)&=\sum_{a=0}^{M} (-1)^a\, p_i(a;c_1,c_2,c_3;N-j)\,
 p_j(y;c_3,c_0,c_4;N-a)\,p_a(x;c_4,c_2,c_1;N-y) \,,\label{eq:G3p}
\end{align}
where  $c_0+c_1+c_2+c_3+c_4=-2N-3$ and $M=\text{min}(N-j,N-x)$.
Due to the properties of the univariate Racah polynomials $p_n(x;c_1,c_2,c_3;N)$, the upper bound $M$ of the sum in \eqref{eq:G3p} can be replaced by $N-x$ or $N-j$.
The notation $G_{i,j}(x,y)$ is short for $G_{i,j}(x,y;c_1,c_2,c_3,c_4;N)$.
As for the Tratnik polynomials, we use the conventions 
$G_{-1,j}(x,y)=G_{i,-1}(x,y)=G_{i,N+1-i}(x,y)=0$. 
Up to a normalization, these polynomials first appeared in \cite{icosi} as the overlap  coefficients between different representations of the rank 2 Racah algebra. 
They can be expressed in terms of the Tratnik polynomials:
\begin{align}
 G_{i,j}(x,y)
 &=\sum_{a=0}^{N-y} (-1)^a\,  
 T_{i,j}(a,y;c_1,c_2,c_3,c_4;N)\,
 p_a(x;c_4,c_2,c_1;N-y)\label{eq:GTp}\\
 & =\sum_{a=0}^{N-j} (-1)^a\,  p_i(a;c_1,c_2,c_3;N-j)\,
  T_{j,a}(y,x;c_3,c_0,c_4,c_1;N)\,.\label{eq:GpT}
\end{align}
In the above expressions, the dependence of the Tratnik polynomials on the parameters $c_j$  is specified.
Substituting the expression \eqref{eq:Tratpol} for $T_{j,a}(y,x;c_3,c_0,c_4,c_1;N)$ 
 into \eqref{eq:GpT},
one gets (for $x,y$ non-negative integers and $x+y\leq N$): 
\begin{align}\label{eq:Gpol}
    G_{i,j}(x,y)&=\Omega(i;c_1,c_2,c_3;N-j)\, (2j+c_{40}+1)\,\frac{(c_3+1)_y}{j!\, (c_0+1)_y}\\
   \times \sum_{a=0}^{N-j}& \frac{(a-N)_j(c_2+1)_a(c_0+1)_{N-a}}{a! (c_{04}+j+1)_{N-a+1} (c_{12}+a+1)_a(c_1+1)_a}(y-N)_a(-N-y-c_{30}-1)_a   \nonumber\\
  &\times \pFq{4}{3}{-i,\;i+c_{23}+1, \;-a,\;a+c_{12}+1}{c_2+1,\; -N-j-1-c_{40},\;j-N}{1}  \pFq{4}{3}{-a,\;a+c_{12}+1, \;-x,\;x+c_{24}+1}{c_2+1,\; -N-y-c_{30}-1,\;y-N}{1}
  \nonumber\\
&\times\pFq{4}{3}{j+a-N,\;N-j+a+c_{123}+2, \;y+a-N,\;a-N-y-c_{30}-1}{2a+c_{12}+2,\;a-c_0-N,\;a-N}{1}\,.
  \nonumber
\end{align}
Writing the hypergeometric functions appearing in \eqref{eq:Gpol} as finite
sums, one can show that $G_{i,j}(x,y)$ divided by the first line of the
r.h.s.~of \eqref{eq:Gpol} 
--- a function of the form $f(i,j)g(x,y)$ --- 
is a bivariate polynomial of total degree $N-j$ w.r.t.~$\lambda(x;c_{24})$ and
$\lambda(y;c_{30})$.

\subsection{Orthogonality}
\begin{prop}[\textbf{Orthogonality}]
The Griffiths polynomials of Racah type satisfy 
the following orthogonality relation:
\begin{align}\label{eq:ortho_griff}
\begin{aligned}
 &\sum_{\genfrac{}{}{0pt}{}{0\leq x,y\leq N}{x+y\leq N}}
 \Lambda(y;c_3,c_0;N)\,\Omega(x;c_1,c_2,c_4;N-y)\,
 G_{i,j}(x,y)\, G_{k,\ell}(x,y)\\
 & \qquad\qquad=\delta_{i,k}\,\delta_{j,\ell}\ 
 \Lambda(j;c_4,c_0;N)\, \Omega(i;c_1,c_2,c_3;N-j)\,.
\end{aligned}
\end{align}
\end{prop}
\proof
Using equation \eqref{eq:LO}, we deduce that, for any $a$,
\begin{equation}\label{eq:Om3}
\frac{\Lambda(y;c_3,c_0;N)}{\Lambda(j;c_4,c_0;N) }=\frac{\Lambda(y;c_3,c_0;N)\,\Lambda(a;c_1,c_2;N)}{\Lambda(j;c_4,c_0;N)\,\Lambda(a;c_1,c_2;N) }=\frac{\Omega(y;c_4,c_0,c_3;N-a)\,\Omega(a;c_3,c_2,c_1;N-j)}{\Omega(j;c_3,c_0,c_4;N-a)\,\Omega(a;c_4,c_2,c_1;N-y)}\,.
\end{equation}
Then, using this property, the expression of the polynomials $G_{i,j}(x,y)$ in terms of the Racah ones \eqref{eq:G3p} as well as the orthogonality relation for the Racah polynomials, the result is proven performing the same calculation as for the Tratnik polynomials.
\endproof

\subsection{Duality}
The Griffiths polynomials of Racah type obey the following duality relation
for $i,j,x,y$ non-negative integers such that $i+j\leq N$ and $x+y\leq N$:
\begin{align}\label{eq:dual-griff}
\frac{G_{i,j}(x,y;c_1,c_2,c_3,c_4;N)}
     {\Omega(i;c_1,c_2,c_3;N-j)\,\Lambda(j;c_4,c_0;N)}=
\frac{G_{x,y}(i,j;c_1,c_2,c_4,c_3;N)}
     {\Omega(x;c_1,c_2,c_4;N-y)\,\Lambda(y;c_3,c_0;N)}\,.
\end{align}
To prove \eqref{eq:dual-griff}, we use the duality relation for the Racah polynomials, as well as \eqref{eq:Om3}.

\subsection{Bispectral property}
The bispectrality of these bivariate polynomials was established in 
\cite{icosi}.
We recall it in the following propositions and provide new, elementary proofs 
of these statements. 

\begin{prop}[\textbf{Recurrence relations}]
\label{prop:rec-griff}
The Griffiths polynomials of Racah type satisfy the following two 
recurrence relations,
for $x,y,i,j$ non-negative integers and $x+y\leq N$, $i+j\leq N$:
\begin{align}\label{eq:rec1-griff}
\Big(\lambda(y;c_{30})+\tfrac{1}{2}(c_3+1)(c_0+1)\Big)\,G_{i,j}(x,y) &=
\sum_{\eps,\eps'=0,\pm1} A^{\eps,\eps'}_{i+\eps,j+\eps'}(c_1,c_2,c_3,c_4)\,
G_{i+\eps,j+\eps'}(x,y)\,,
\end{align}
where the coefficients $A^{\eps,\eps'}_{i,j}(c_1,c_2,c_3,c_4)$ are given 
in \eqref{def:Aij}, 
and
\begin{subequations}\label{def:Gam}
\begin{align}\label{eq:rec2-griff}
\begin{aligned}
&\Big(\lambda(x;c_{42})+\tfrac{1}{2}(c_2+1)(c_4+1)\Big)\,G_{i,j}(x,y)\\
&\qquad\qquad
=\sum_{\eps,\eps'=0,\pm1}\left(
 A^{\eps,\eps'}_{i+\eps,j+\eps'}(c_1,c_2,c_3,c_4)-
 \Gamma^{\eps,\eps'}_{i+\eps,j+\eps'} \right)\,G_{i+\eps,j+\eps'}(x,y)\,,
\end{aligned}
\end{align}
where 
\begin{gather}
\Gamma_{i,j}^{+,+} \ =\ \Gamma_{i,j}^{-,+}\ =\ \Gamma_{i,j}^{+,-}\ 
 =\ \Gamma_{i,j}^{-,-}\ =\ 0\,,\\
\Gamma_{i,j}^{0,+} =C_j(c_1,c_0,c_4;N-i)\,,\qquad 
\Gamma_{i,j}^{0,-} =A_j(c_1,c_0,c_4;N-i)\,,\\
\Gamma_{i,j}^{+,0} =C_{i}(c_1,c_2,c_3;N-j)\,,\qquad 
\Gamma_{i,j}^{-,0} = A_{i}(c_1,c_2,c_3;N-j)\,,
\end{gather}
and
\begin{align}
\Gamma_{i,j}^{0,0} 
&\equiv \Gamma_{i,j}^{0,0}(c_1,c_2,c_3,c_4)=
-\Sigma_{i}(c_1,c_2,c_3;N-j)+\Sigma_{j}(c_1,c_0,c_4;N-i)
-\tfrac{1}{4}(c_3^2-c_4^2)\nonumber\\
&\quad+\left(i-N-\tfrac{1}{2}(c_4+1)\right)
       \left(i+\tfrac{1}{2}(c_{23}-c_{01})\right)
 -\left(j-N-\tfrac{1}{2}(c_3+1)\right)
  \left(j+\tfrac{1}{2}(c_{04}-c_{12})\right)\,.
\end{align}
\end{subequations}
\end{prop}
\proof
The proof of \eqref{eq:rec1-griff} proceeds by writing the polynomials 
$G_{i,j}(x,y)$
in the form \eqref{eq:GTp} and using the recurrence relation \eqref{def:Aij} of the Tratnik polynomials.

To prove \eqref{def:Gam}, we rewrite the l.h.s.~of \eqref{eq:rec2-griff} 
using the expression 
\eqref{eq:GpT} for the Griffiths polynomials of Racah type and 
use the recurrence relation \eqref{def:Aij} to get
\begin{align}\label{eq:delicat}
 &\sum_{a=0}^{N-j} \sum_{\epsilon,\epsilon'=0,\pm 1}(-1)^a p_i(a;c_1,c_2,c_3;N-j)\, A^{\epsilon,\epsilon'}_{j+\epsilon,a+\epsilon'}(c_3,c_0,c_4,c_1)\,T_{j+\epsilon,a+\epsilon'}(y,x;c_3,c_0,c_4,c_1;N)\\
 \hspace{-3em}=&\sum_{\epsilon,\epsilon'=0,\pm 1} 
 \sum_{a=\epsilon'}^{{N-j}+\epsilon'}
 (-1)^{a-\epsilon'} p_i(a-\epsilon';c_1,c_2,c_3;N-j) \,
 A^{\epsilon,\epsilon'}_{j+\epsilon,a}(c_3,c_0,c_4,c_1)\,
 T_{j+\epsilon,a}(y,x;c_3,c_0,c_4,c_1;N)\,.\nonumber
\end{align}
The r.h.s. of \eqref{eq:delicat} can be transformed as
\begin{align}\label{eq:delicat2}
 \sum_{\epsilon=0,\pm 1} 
 \sum_{a=0}^{{N-j}-\epsilon}
\sum_{\epsilon'=0,\pm 1} 
 (-1)^{a-\epsilon'} p_i(a-\epsilon';c_1,c_2,c_3;N-j) \,
 A^{\epsilon,\epsilon'}_{j+\epsilon,a}(c_3,c_0,c_4,c_1)\,
 T_{j+\epsilon,a}(y,x;c_3,c_0,c_4,c_1;N)\,.
\end{align}
To get to this formula, the boundaries of the $a$ summation
in the second line of \eqref{eq:delicat} have been changed.
It is easily checked that each of the boundary terms that have been added
or removed are identically zero and thus have no contribution.
For instance, when $\epsilon'=-1$, we can discard the contribution of $a=-1$ 
since $T_{j+\epsilon,-1}(y,x;c_3,c_0,c_4,c_1;N)=0$. 
Similarly, when $\epsilon'=1$, we can add the term $a=0$ to the sum because 
$A^{\epsilon,{+}}_{j+\epsilon,0}=0$. 

Next, one needs  the following relations, proven in Appendix \ref{App:A}, that are  satisfied 
by the Racah polynomial for $\epsilon=0,\pm1$ 
and $0\leq a\leq N-j-\epsilon$:
\begin{align}\label{eq:Le1}
 &\sum_{\epsilon'=0,\pm 1} (-1)^{\epsilon'}\,p_i(a-\epsilon';c_1,c_2,c_3;N-j)\,A_{j+\epsilon,a}^{\epsilon,\epsilon'}(c_3,c_0,c_4,c_1)\nonumber\\
 &\qquad=\sum_{\mu=0,\pm 1} p_{i+\mu}(a;c_1,c_2,c_3;N-j-\epsilon) \,\left(A_{i+\mu,j+\epsilon}^{\mu,\epsilon}(c_1,c_2,c_3,c_4)-\Gamma_{i+\mu,j+\epsilon}^{\mu,\epsilon}\right)\,,
\end{align}
where the $\Gamma^{\mu,\epsilon}_{i,j}$ are given in \eqref{def:Gam}. 
Finally, we recognize in \eqref{eq:delicat2} the l.h.s.~of relation  
\eqref{eq:Le1}, to conclude the proof.
\endproof

\begin{prop}[\textbf{Difference equations}]
The Griffiths polynomials of Racah type,
for $x,y,i,j$ non-negative integers and $x+y\leq N$, $i+j\leq N$ obey the following two 
difference equations:
\begin{align}\label{eq:diff1-griff}
\Big(\mu_j(c_{04})+\tfrac{1}{2}(c_4+1)(c_0+1)\Big)\,G_{i,j}(x,y) 
=\sum_{\eps,\eps'=0,\pm1}D^{\eps',\eps}(y,x;c_3,c_0,c_4,c_1)\,
 G_{i,j}(x+\eps,y+\eps')\,,
\end{align}
where the coefficients $D^{\eps,\eps'}(x,y;c_1,c_2,c_3,c_4)$ are given 
in \eqref{def:Dij}, and
\begin{subequations}\label{def:Psi}
\begin{align}\label{eq:diff2-griff}
\begin{aligned}
&\Big(\mu_i(c_{23})+\tfrac{1}{2}(c_2+1)(c_3+1)\Big)\,G_{i,j}(x,y)\ \\
&\qquad=\sum_{\eps,\eps'=0,\pm1}\Big(D^{\eps',\eps}(y,x;c_3,c_0,c_4,c_1)-
\Psi^{\eps',\eps}(x,y)\Big)\,G_{i,j}(x+\eps,y+\eps')\,,\quad
\end{aligned}
\end{align}
with 
\begin{gather}
\Psi^{+,+}(x,y) \ =\ \Psi^{-,+}(x,y)\ =\ \Psi^{+,-}(x,y)\ 
=\ \Psi^{-,-}(x,y)\ =\ 0\,,\\
\Psi^{0,+}(x,y) =B(x;c_4,c_2,c_1,N-y)\,,\qquad 
\Psi^{0,-}(x,y) =D(x;c_4,c_2,c_1,N-y)\,,\\
\Psi^{+,0}(x,y) =B(y;c_3,c_0,c_1,N-x)\,,\qquad 
\Psi^{-,0}(x,y) = D(y;c_3,c_0,c_1,N-x)\,, \\
\Psi^{0,0}(x,y;c_3,c_0,c_4,c_1) =
\Gamma_{x,y}^{0,0}(c_1,c_2,c_4,c_3)\,.
\end{gather}
\end{subequations}
\end{prop}
\proof 
The proof can be done  mimicking the one of Proposition \ref{prop:rec-griff}.
Alternatively, one can use the duality property \eqref{eq:dual-griff} of 
the Griffiths polynomials of Racah type to 
obtain the result directly from the recurrence relations. 
\endproof

\subsection{Domains\label{sect:domains}}
Up to now, the pairs $(i,j)$ and the variables $(x,y)$ in the recurrence relations
and difference equations of the Griffiths polynomials $G_{i,j}(x,y)$ have been
restrained to non-negative integers such that $i+j \leq N$ and $x+y \leq N$. 

When a single parameter $c_i$ is specialized to some negative integer values, the subspaces of $\mathbb{Z}_{\geq 0}^2$ in which $(i,j)$ and $(x,y)$ take values respectively might have to be restrained further. We will refer to these subspaces of $\mathbb{Z}_{\geq 0}^2$ as domains for $(i,j)$ and $(x,y)$. The polynomials $G_{i,j}(x,y)$ can be defined on these restricted domains. Their recurrence relations and difference equations read as \eqref{eq:rec1-griff}, \eqref{def:Gam}, \eqref{eq:diff1-griff} and \eqref{def:Psi} for $(i,j)$ and $(x,y)$ inside the domains, with the convention that the polynomials $G_{i,j}(x,y)$ involving pairs $(i,j)$ or $(x,y)$ outside the domains and which appear in these equations are 
consistently set  to zero.

\begin{prop} \label{prop:domains} 
Let $k \in \{1,2,\dots,N\}$. For each specialization listed below, one can consider a restricted set of polynomials $G_{i,j}(x,y)$ which satisfy the recurrence relations \eqref{eq:rec1-griff}--\eqref{def:Gam} and the difference equations \eqref{eq:diff1-griff}--\eqref{def:Psi} on restricted domains for $(i,j)$ and $(x,y)$. In all cases, $i,j,x,y$ are non-negative integers such that $i+j \leq N$, $x+y\leq N$, and
the domains for $(i,j)$ and $(x,y)$ defined by:
\begin{align}
&\label{eq:dom0} \bullet\ \text{if } c_0=-k, \left\{\begin{array}{l l l}
    j< k, &  y< k, &\quad \text{with } G_{i,k}(x, y)=0,\\
    \qquad\text{or}\\
    j\geq k, & y\geq k, & \quad\text{with } G_{i,j}(x,k-1)=0;
\end{array}  \right.\\
 &\label{eq:dom1}\bullet\ \text{if } c_1=-k, \left\{\begin{array}{l l l}
  i+j > N-k,&   x+y > N-k,& \quad \text{with } G_{i,N-k-i}(x,y)=0,\\
    \qquad\text{or}\\
   i+j \leq  N-k,&   x+y \leq  N-k,& \quad \text{with } G_{i,j}(x,N-k-x+1)=0;
\end{array}  \right.\\
 &\label{eq:dom2}\bullet\ \text{if } c_2=-k, \left\{\begin{array}{l l l}
  i< k, & x< k,& \quad \text{with } G_{k,j}(x, y)=0,\\
    \qquad\text{or}\\
   i\geq k,&  x\geq k,& \quad \text{with } G_{i,j}(k-1,y)=0;
\end{array}  \right.\\
 &\label{eq:dom3}\bullet\ \text{if } c_3=-k, \left\{\begin{array}{l l l}
i< k,&  y< k, &\quad \text{with } G_{i,j}(x,k)=0,\\
    \qquad\text{or}\\
   i\geq k,&  y\geq k,& \quad \text{with } G_{k-1,j}(x, y)=0;
\end{array}  \right.\\
&\label{eq:dom4}\bullet\ \text{if } c_4=-k, \left\{\begin{array}{l l l}
j< k,&   x< k, &\quad \text{with } G_{i,j}(k,y)=0,\\
    \qquad\text{or}\\
 j\geq k, &  x\geq k, &\quad \text{with } G_{i,k-1}(x,y)=0.
\end{array}  \right.
\end{align}
\end{prop}
\proof 
Let us examine in detail the case $c_2=-k$. 

First note that although there might be singularities in some of the 
hypergeometric functions appearing in $G_{i,j}(x,y)$, the limit is still 
well-defined by virtue of the normalization coefficients $\Omega$. 
Using the explicit expression \eqref{eq:G3p}, it can also be verified that 
$G_{i,j}(x,y)=0$ if $i=k,k+1,\dots,N$ and $x=0,1,\dots,k-1$. 
Indeed, if $a<k$ in the sum of \eqref{eq:G3p}, there is no singularity 
in $p_i(a;c_1,c_2,c_3;N-j)$ but there is a coefficient $(c_2+1)_i$ which is
zero since $i\geq k$. Similarly, if $a \geq k$ and since $x < k$, there is 
no singularity in $p_a(x;c_4,c_2,c_1;N-y)$ but there is a coefficient 
$(c_2+1)_a$ which is zero.   

Second, by examining the explicit expressions of the recurrence coefficients in \eqref{eq:rec1-griff} and \eqref{def:Gam}, one finds that $A^{-,\varepsilon'}_{k-1,j+\varepsilon'}=0$ and $\Gamma^{-,\varepsilon'}_{k-1,j+\varepsilon'}=0$ for $\varepsilon'=0,\pm 1$ and any $j$. It follows from this that all polynomials $G_{i,j}(x,y)$ with $i \geq k$ can be constructed from $G_{k,N-k}(x,y)$ using the recurrence relations, independently from those with $i < k$. 

Third, the coefficients in the difference equations \eqref{eq:diff1-griff} and \eqref{def:Psi} are such that $D^{\varepsilon',+}(y,k-1)=0$ and $\Psi^{\varepsilon',+}(k-1,y)=0$ for $\varepsilon'=0,\pm 1$ and any $y$.

By combining all previous observations, one deduces that there are two possible ways of restraining the domains for $(i,j)$ and $(x,y)$. The first option is to take $i,x<k$. Indeed, in this case one needs to impose $G_{i,j}(x,y)=0$ for $i\geq k$, since these polynomials might appear as terms with non-vanishing coefficients in the recurrence relations. Then, in order for the recurrence relations to hold with this convention for all pairs $(i,j)$ in their restricted domain, one must restrict $x<k$, as the polynomials $G_{i,j}(x,y)$ are zero for $i\geq k$ and $x<k$. In the difference equations, terms containing polynomials $G_{i,j}(x,y)$ with $x\geq k$ do not appear because of the vanishing coefficients. The second option is to take $i,x \geq k$. In this case, the previous arguments concerning the recurrence relations and the difference equations are reversed, and one should impose $G_{i,j}(x,y)=0$ for $x<k$.  

A similar analysis applies for the other parameters. We indicate below 
the vanishing of the polynomials and of the recurrence/difference 
coefficients. \par\vspace{1ex}

When $c_0=-k$, one finds that $G_{i,j}(x,y)=0$ for $j\geq k$ and $y<k$ using the polynomials $p_j(y;c_3,c_0,c_4;N-a)$. The coefficients $A^{\varepsilon,-}_{i+\varepsilon,k-1}$, $\Gamma^{\varepsilon,-}_{i+\varepsilon,k-1}$, $D^{+,\varepsilon}(k-1,x)$ and $\Psi^{+,\varepsilon}(x,k-1)$ vanish.  

When $c_3=-k$, one can use Whipple's transformation formula on $p_i(a;c_1,c_2,c_3;N-j)$ and $p_j(y;c_3,c_0,c_4;N-a)$, see equation (2.10.5) in \cite{GR}, to show that $G_{i,j}(x,y)=0$ for $i<k$ and $y \geq k$. The coefficients $A^{+,\varepsilon'}_{k,j+\varepsilon'}$, $\Gamma^{+,\varepsilon'}_{k,j+\varepsilon'}$, $D^{-,\varepsilon}(k,x)$ and $\Psi^{-,\varepsilon}(x,k)$ vanish.  

When $c_4=-k$, one can use again Whipple's transformation formula on $p_j(y;c_3,c_0,c_4;N-a)$ and $p_a(x;c_4,c_2,c_1;N-y)$ to show that $G_{i,j}(x,y)=0$ for $j<k$ and $x\geq k$. The coefficients $A^{\varepsilon,+}_{i+\varepsilon,k}$, $\Gamma^{\varepsilon,+}_{i+\varepsilon,k}$, $D^{\varepsilon',-}(y,k)$ and $\Psi^{\varepsilon',-}(k,y)$ vanish. 

Finally, when $c_1=-k$, applying Whipple's transformation formula on $p_i(a;c_1,c_2,c_3;N-j)$ and $p_a(x;c_4,c_2,c_1;N-y)$ shows that $G_{i,j}(x,y)=0$ for $i+j \leq N-k$ and $x+y > N-k$. The coefficients $A^{\varepsilon,\varepsilon'}_{i+\varepsilon,j+\varepsilon'}$, $\Gamma^{\varepsilon,\varepsilon'}_{i+\varepsilon,j+\varepsilon'}$ vanish for the following values: $(\varepsilon,\varepsilon')=(+,0),(0,+)$ and $i+j=N-k$, $(\varepsilon,\varepsilon')=(+,+)$ and $i+j=N-k,N-k-1$. Moreover, the coefficients $D^{\varepsilon',\varepsilon}(y,x)$ and $\Psi^{\varepsilon',\varepsilon}(x,y)$ vanish for: $(\varepsilon,\varepsilon')=(-,0),(0,-)$ and $x+y=N-k+1$, $(\varepsilon,\varepsilon')=(-,-)$ and $x+y=N-k+1,N-k+2$.
\endproof

Proposition \ref{prop:domains} indicates that the recurrence relations and 
difference equations still hold on a restricted domain 
even if a parameter $c_i$ is a negative integer.  
The following proposition states that these polynomials are also 
orthogonal on the corresponding domains. 
\begin{prop}
Consider the specializations listed in Proposition \ref{prop:domains}, and denote respectively by $\cD_1$ and $\cD_2$ the restricted domains of $(i,j)$ and $(x,y)$ given in \eqref{eq:dom0}--\eqref{eq:dom4}. For each choice of domains, the restricted set of polynomials $G_{i,j}(x,y)$ with $(i,j) \in \cD_1$ satisfy the following orthogonality relation:
\begin{align}\label{eq:ortho2_griff}
&\sum_{(x,y) \in \cD_2} \widetilde{\Lambda}(y;c_3,c_0;N)\,
 \widetilde{\Omega}(x;c_1,c_2,c_4;N-y)\,
 G_{i,j}(x,y)\, G_{m,n}(x,y)\nonumber\\
 & \qquad\qquad   =\delta_{i,m}\,\delta_{j,n}\ \overline{\Lambda}(j;c_4,c_0;N)\, \overline{\Omega}(i;c_1,c_2,c_3;N-j)\,,
\end{align}
where
$\widetilde{\Omega}, \overline{\Omega}$ and $\widetilde{\Lambda}, \overline{\Lambda}$ are non-zero functions deduced from 
$\Omega$ and $\Lambda$ in equations \eqref{def:Omega} and \eqref{def:Lambda} through the relations
\begin{equation}\label{eq:Omega2}
\begin{split}
&\frac{\widetilde\Lambda(y;c_3,c_0;N)\,\widetilde\Omega(x;c_1,c_2,c_4;N-y)}{\overline\Lambda(j;c_4,c_0;N)\,\overline\Omega(i;c_1,c_2,c_3;N-j)}=  \lim_{c_\ell\to-k}\frac{\Lambda(y;c_3,c_0;N)\,\Omega(x;c_1,c_2,c_4;N-y)}{\Lambda(j;c_4,c_0;N)\,\Omega(i;c_1,c_2,c_3;N-j)}\,.
\end{split}
\end{equation}
\end{prop}
\proof
Let us examine again in detail the case $c_2=-k$. The orthogonality relation \eqref{eq:ortho_griff} still holds for the polynomials $G_{i,j}(x,y)$. It is seen from \eqref{def:Omega} that $\Omega(n;c_1,-k,c_3;N)=0$ for $n\geq k$, but otherwise for $n<k$ this function does not vanish in general. It follows that for the set of polynomials $G_{i,j}(x,y)$ restricted by the condition $i<k$, the sum in \eqref{eq:ortho_griff} can be restrained to the pairs $(x,y)$ which satisfy $x<k$. This yields the restrained orthogonality relation in the upper line of \eqref{eq:dom2}. Equation \eqref{eq:ortho_griff} becomes trivial for the set of polynomials $G_{i,j}(x,y)$ restricted by $i\geq k$. Indeed, on the r.h.s.~the function $\Omega$ vanishes and on the l.h.s.~we have either that $G_{i,j}(x,y)=G_{m,n}(x,y)=0$ for $x<k$ (see proof of Proposition \ref{prop:domains}) or that the function $\Omega$ vanishes for $x\geq k$. To obtain a non-trivial relation, one can restrain the sum to pairs $(x,y)$ such that $x\geq k$ and simplify the vanishing factors $(c_2+k)$ appearing in the functions $\Omega$ both on the l.h.s.~and on the r.h.s. This gives an orthogonality relation for the lower line of \eqref{eq:dom2} with modified functions $\widetilde{\Omega}=\overline{\Omega}$ as in \eqref{eq:Omega2}. 

All other cases can be similarly treated by restraining the sum in \eqref{eq:ortho_griff} and by simplifying vanishing factors in the functions $\Omega$ and $\Lambda$, when necessary, in order to obtain well-defined and non-trivial orthogonality relations.
\endproof

We have examined in this subsection restrictions on the domains when only one parameter is specialized. It is possible to consider other situations, for instance a combination of the specializations listed in Proposition \ref{prop:domains}, where more than one parameter takes a negative integer value in $\{-1,-2,\dots,-N\}$ (not necessarily the same value for all specialized parameters). 
Restrictions on the domains can be determined using similar arguments as the ones in the proof of Proposition \ref{prop:domains}, but 
each situation requires its own analysis.  

\subsection{Connection with \texorpdfstring{$9j$}{9j} symbols}
An example of a situation where the parameters $\{c_i\}_{i=0}^4$
simultaneously take negative integer values arises in the study of 
the $9j$ symbols.

%

\begin{prop}
For $c_1$, $c_2$, $c_3$, $c_4$ and $c_0$ negative integers obeying 
the constraint \eqref{eq:contrainte}, and assuming that 
some additional constraints that are made explicit in the core of the proof
are obeyed, 
the Griffiths polynomials of Racah type \eqref{eq:G3p} 
are proportional to the $9j$ symbol
\begin{align}\label{eq:G9j}
 G_{i,j}(x,y;c_1,c_2,c_3,c_4)\propto
 \left\{
 \begin{matrix}
  \phantom{-i}\!-\!\frac{1}{2}(c_2\!+\!1) & 
  \phantom{-j}\!-\!\frac{1}{2}(c_4\!+\!1) & 
  -x-\frac{1}{2}(c_{24}\!+\!2) 
  \\[.2em]
  \phantom{-i}\!-\!\frac{1}{2}(c_3\!+\!1) & 
  \phantom{-j}\!-\!\frac{1}{2}(c_0\!+\!1) & 
  -y-\frac{1}{2}(c_{03}\!+\!2) 
  \\[.2em]
  -i\!-\!\frac{1}{2}(c_{23}\!+\!2) & 
  -j\!-\!\frac{1}{2}(c_{04}\!+\!2) & 
  \phantom{-x!}\!-\!\frac{1}{2}(c_1\!+\!1)
 \end{matrix}
 \right\}\,.
\end{align} 
The symbol $\propto$ in the above relation means 
equality up to a non-zero function of type $f(i,j)g(x,y)$ 
which can be computed explicitly.  
\end{prop}

\proof
The $9j$ symbols can be expressed in terms of $6j$ symbols
\cite{KMV}
\begin{align}\label{eq:9j6j} 
\left\{
 \begin{matrix}
  j_1 & j_2 & j_{12} \\
  j_3 & j_4 & j_{34} \\
  j_{13} & j_{24} & j_{0}
 \end{matrix}
 \right\}
 =\sum_{j_{234}}(-1)^{2j_{234}}(2j_{234}+1)
 \left\{
 \begin{matrix}
  j_{24} & j_3 & j_{234} \\
  j_1 & j_{0} & j_{13} 
 \end{matrix}
 \right\}
 \left\{
 \begin{matrix}
  j_{234} & j_2 & j_{34} \\
  j_4 & j_3 & j_{24} 
 \end{matrix}
 \right\}
 \left\{
 \begin{matrix}
  j_{34} & j_{0} & j_{12} \\
  j_1 & j_2 & j_{234} 
 \end{matrix}
 \right\}\,,
\end{align}
where we have used symmetries of the $6j$ symbol to rewrite them in 
a convenient way for what follows.
Note that the $6j$ symbol is equal to zero unless the so-called triangular 
conditions are obeyed \cite{KMV}.
Moreover, $6j$ symbols are related to ${}_{4}F_{3}$ generalized 
hypergeometric functions by
\begin{align}
\left\{
 \begin{matrix}
  j_{123} & j_{1} & j_{23} \\
  j_2 & j_3 & j_{12} 
 \end{matrix}
\right\}=
\mathcal{N}_{j_2,j_3,j_{12}}^{j_{123},j_1,j_{23}}\,
 \pFq{4}{3}
 {j_{12}\!-\!j_1\!-\!j_2,\;-j_{12}\!-\!j_1\!-\!j_2\!-\!1,\;
  j_{23}\!-\!j_2\!-\!j_3,\;-j_{23}\!-\!j_2\!-\!j_3\!-\!1}
 {-2j_2,\;j_{123}\!-\!j_1\!-\!j_2\!-\!j_3,\;
  -j_{123}\!-\!j_1\!-\!j_2\!-\!j_3\!-\!1}{1}\,
\end{align}
when the following additional constraints\footnote{
Other connections with ${}_{4}F_{3}$ functions can be found, 
depending on the type of additional inequalities one chooses, 
see relations (8--12) of Chapter 9.2.3 in \cite{KMV}
and Chapter II of \cite{Wilson} for more details on these relations.
} on the entries of the $6j$ symbol are obeyed \cite{Wilson}:
\begin{align}\label{eq:Wilson_constraints}
\begin{aligned}
 j_{123}+j_1&\geq j_2+j_3 \quad\text{and}\quad j_{123}-j_1\geq|j_2-j_3|\,. 
\end{aligned}
\end{align}
The normalization is given by
\begin{align}
\begin{aligned}
\mathcal{N}_{j_2,j_3,j_{12}}^{j_{123},j_1,j_{23}}&=
 (-1)^{j_1+j_2+j_3+j_{123}}(2j_2)!\,(j_1\!+\!j_2\!+\!j_3\!-\!j_{123})!\,
 (j_1\!+\!j_2\!+\!j_3\!+\!j_{123}\!+\!1)!\\
 &\quad\times\Delta(j_1,j_2,j_{12})\,\Delta(j_{12},j_3,j_{123})\,
 \Delta(j_{23},j_2,j_3)\,\Delta(j_{123},j_1,j_{23})\,,
\end{aligned}
\end{align}
and the symbol $\Delta(j_1,j_2,j_3)$ is defined as
\begin{align}
 \Delta(j_1,j_2,j_3)=\sqrt{\frac{(j_1\!-\!j_2\!+\!j_3)!}
  {(-j_1\!+\!j_2\!+\!j_3)!\,(j_1\!+\!j_2\!+\!j_3\!+\!1)!\,
   (j_1\!+\!j_2\!-\!j_3)!}}\,.
\end{align}
Assuming that the constraints \eqref{eq:Wilson_constraints} hold 
for each of the $6j$ symbols appearing in \eqref{eq:9j6j}
and replacing the summation over $j_{234}$ by a summation over the variable
$a=-j_{234}-\frac{1}{2}(2+c_{12})$,
we get the following constraints between the parameters $\{c_i\}_{i=0}^4$,
the degree $j$, variable $y$ and summation index $a$:
\begin{equation}\label{eq:9j-cont2}
\begin{aligned}
 -j+N+1+c_{12}\geq0\,,\qquad&
 -2j-1-c_{04}+c_3\geq |c_1-c_2|\,,\\
 -a+N+1+c_{03}\geq0\,,\qquad&
 -2a-1-c_{12}+c_4\geq |c_0-c_3|\,,\\
 -y+N+1+c_{24}\geq0\,,\qquad &
 -2y-1-c_{03}+c_1\geq |c_2-c_4|\,.
\end{aligned}
\end{equation}
Then, using the relation between $6j$ symbols and 
hypergeometric functions above,
we can show that the r.h.s.~of \eqref{eq:G9j} 
becomes the desired Griffiths polynomial of Racah type
multiplied by a function of type $f(i,j)g(x,y)$ which we do not 
write here.
Note that upon setting $c_j$'s to negative integers,
zeros appear in both the numerator and denominator of 
the proportionality coefficient of \eqref{eq:G9j} 
but closer inspection shows that
all zeros cancel under the limiting procedure.
\endproof

\section{Limiting cases of Griffiths polynomials of Racah type}
\label{sect:limit}

We now look for limits $|c_j|\to\infty$ for some of the parameters of 
the Griffiths polynomials of Racah type while preserving the relation 
\eqref{eq:contrainte}.
This implies that at least 2 $c$'s must tend to $\pm\infty$. 
After examining all possible combinations where
$2$, $3$, $4$ or $5$ $c$'s tend to infinity at the same speed 
(\textit{i.e.}~$c_j=\sigma_j\,t$ with $t\to\infty$), we find
various cases of well-defined limits of the Griffiths polynomials
$G_{i,j}(x,y)$, presented below.
The Griffiths polynomials of Racah type are orthogonal 
and bispectral, and we checked that the same properties 
also hold for the limiting cases. 
The proof is done by considering the limit of relations 
\eqref{eq:ortho_griff}, \eqref{eq:rec1-griff}--\eqref{def:Gam} and 
\eqref{eq:diff1-griff}--\eqref{def:Psi}, 
possibly multiplied by an appropriate power of $t$.
Surprisingly, only two types of limits lead to non-trivial, well-defined polynomials: when two or when all $c$'s tend to infinity.

To present the corresponding polynomials,
we first recall the definition of the Hahn polynomials 
$H_n(x;c_1,c_2,N)$ and dual Hahn polynomials $\widetilde{H}_n(x;c_1,c_2,N)$
\cite{Koek}
\begin{align}
 H_{n}(x;c_{1},c_{2};N) &= 
\binom{N}{n}\,(2n+c_{1}+1)\frac{(c_2+1)_n}{(c_1+n+1)_{N+1}}\ 
\pFq{3}{2}{-x,\;-n,\;n+c_{1}+1}{c_{2}+1,\;-N}{1}\,,\\
\widetilde H_{n}(x;c_{1},c_{2};N) &= \binom{N}{n}\,(c_2+1)_{n}\ 
\pFq{3}{2}{-n,\;-x,\;x+c_{1}+1}{c_2+1,\;-N}{1}\,,
\end{align}
as well as the Krawtchouk polynomial \cite{Koek}
\begin{equation}\label{eq:kraw}
K_n(x;\fp;N)=  \binom{N}{n} \left( \frac{\fp}{1-\fp}\right)^n 
\pFq{2}{1}{-n,\;-x}{-N}{\frac{1}{\fp}}\,.
\end{equation}
\subsection{Two \texorpdfstring{$c$}{c}'s tend to infinity}
Let us first consider a renormalized version of the Griffiths polynomials \eqref{eq:G3p}:
\begin{equation}
\tilde G_{i,j}(x,y)=\frac{(c_3+1)_{N-j}}{(c_4+1)_{N-y}}\sum_{a=0}^{N-j} (-1)^a\, p_i(a;c_1,c_2,c_3;N-j)\,p_j(y;c_3,c_0,c_4;N-a)\,p_a(x;c_4,c_2,c_1;N-y) \,.\label{eq:G3p_norm}
\end{equation}

\paragraph{Griffiths polynomials of type dual Hahn -- dual Hahn -- Racah.} 
Shift the parameters $c_0,c_3$ according to 
$c_0\mapsto c_0-t$, $c_3\mapsto c_3+t$ 
and perform the limit $t\to\infty$.
The polynomials $\tilde{G}_{i,j}(x,y)$ tend to
\begin{equation}\label{eq:griff_dHdHR}
\begin{split}
\tilde h_{i,j}(x,y)=&\sum_{a=0}^{N-j} (-1)^{N+j}\,\frac{(c_1+1)_{N-j-i}}{(c_4+1)_{N-y}\,(c_4+1)_{j}}\ \widetilde H_i(a;c_{12},c_{2};N-j)\,\\
&\qquad\times \widetilde H_j(y;c_{30},c_{304}+N-a+1;N-a)\,
p_a(x;c_4,c_2,c_1;N-y) \,.
\end{split}
\end{equation}
\paragraph{Griffiths polynomials of type Racah -- Hahn -- Hahn.} 
Shift the parameters $c_0,c_4$ according to 
$c_0\mapsto c_0-t$, $c_4\mapsto c_4+t$ 
and take the limit $t\to\infty$.
The polynomials $\tilde{G}_{i,j}(x,y)$ tend to
\begin{equation}\label{eq:griff_RHH}
\begin{split}
h_{i,j}(x,y)=&\sum_{a=0}^{N-j} (-1)^{a+j}\,\frac{(c_3+1)_{N-j-a}\,(c_3+1)_{N-j}}{(c_1+1)_{a}}\ 
p_i(a;c_1,c_2,c_3;N-j)\, \\
&\qquad\times H_j(y;c_{04},c_{304}+N-a+1;N-a)\,H_a(x;c_{12},c_{2};N-y) \,.
\end{split}
\end{equation}

\paragraph{Griffiths polynomials of type dual Hahn -- Racah -- Hahn.} 
Shift the parameters $c_1,c_2$ according to 
$c_1\mapsto c_1+t$, $c_2\mapsto c_2-t$ 
and perform the limit $t\to\infty$.
The polynomials $\tilde{G}_{i,j}(x,y)$ tend to
\begin{equation}\label{eq:griff_dHRH}
\begin{split}
\bar h_{i,j}(x,y)=&\sum_{a=0}^{N-j} (-1)^{N+i+j}\, 
\frac{(c_3+1)_{N-j}\,(c_4+1)_{N-y-a}}{(c_4+1)_{N-y}\,(c_3+1)_{i}}\
\widetilde H_i(a;c_{12},c_{123}+N-j+1;N-j)\, \\
&\qquad\times p_j(y;c_3,c_0,c_4;N-a)\,H_a(x;c_{12},c_{124}+N-y+1;N-y) \,.
\end{split}
\end{equation}

\subsection{All \texorpdfstring{$c$}{x}'s tend to infinity: 
Griffiths polynomials (of Krawtchouk type)}

Suppose now that all the parameters  $c_i$ scale as $\sigma_i t$ in 
 the limit  $t\to \infty$. 
Once again, assume that the constraint \eqref{eq:contrainte} is fulfilled, 
which implies that 
$\sigma_0+\sigma_1+\sigma_2+\sigma_3+\sigma_4=0$. 
This limiting procedure relates the Racah polynomials to 
the Krawtchouk polynomials \eqref{eq:kraw}:
\begin{align}
\lim_{t\to\infty}p_n(x;t\sigma_i,t\sigma_j,t\sigma_k;N)= 
\left(\frac{\sigma_i}{\sigma_j+\sigma_k}\right)^N\! 
K_n(x;\fp_{ijk};N)\,,\qquad
\fp_{ijk}=\frac{\sigma_j(\sigma_i+\sigma_j+\sigma_k)}
               {(\sigma_i+\sigma_j)(\sigma_j+\sigma_k)}
\,.
\end{align}
Under the above parametrization, one obtains
\begin{align}
\lim_{t\to\infty}G_{i,j}(x,y)&= \left(\frac{\sigma_1}{\sigma_2+\sigma_3}\right)^{N-j}\,\left(\frac{\sigma_3}{\sigma_0+\sigma_4}\right)^{N}\,\left(\frac{\sigma_4}{\sigma_1+\sigma_2}\right)^{N-y}\,\tilde g_{i,j}(x,y) \,,\\
\tilde g_{i,j}(x,y)&= 
\sum_{a=0}^{N-j} \left(-\frac{\sigma_0+\sigma_4}{\sigma_3}\right)^a\,
K_i(a;\fp_{123};N-j)\, K_j(y;\fp_{304};N-a)\, K_a(x;\fp_{421};N-y)\,.
\label{eq:griff_kkk}
\end{align}
The polynomials $\tilde g_{i,j}(x,y)$ are proportional to the Griffiths polynomials (of Krawtchouk type) introduced in \cite{Griff}.

\section{Outlook}
In this paper, we have presented a straightforward characterization 
of the bivariate Griffiths polynomials of Racah type introduced in \cite{icosi}. 
The approach should be extendable to 
multivariate polynomials.

The methods introduced here open the path to various further studies. 
For instance, in Section~\ref{sect:domains}, specialization to negative integers for 
the individual $c_j$'s were performed to get restricted domains.
The same kind of specialization could also be possible when considering sums 
of the form $c_j+c_k$. 
Moreover, in the limiting procedure studied in Section \ref{sect:limit},
we did not investigate the case when the speeds at which the parameters $c_j$'s go to infinity are different. 
This could lead to novel types of Griffiths polynomials. 

One may also wonder if the approach  works in 
the $q$-deformed case. Indeed, the Tratnik polynomials of $q$-Racah type are
known and have been studied \cite{GR07,Il2011}. However, the Griffiths polynomials of 
$q$-Racah type have not been discovered yet. 
We believe that the methods developed in the present paper could inform this problem. Recall that the Griffiths polynomials of Racah type have been obtained in \cite{icosi} through the construction of the representation of the  rank 2 Racah algebra (see also \cite{SP2024} for a recent study of these representations). It is to be expected that constructing analogously the representations of the rank 2 Askey--Wilson algebra\footnote{See \cite{PW, DDv, awn} for the definition of higher rank Askey--Wilson algebras.} will lead to the identification of the bivariate Griffiths polynomials of $q$-Racah type.  Eventually, 
the limit from these functions to $q$-Krawtchouk polynomials of two variables
should provide a realization of the $SU_q(3)$ quantum group representation basis, 
as already seen in \cite{Bergeron} for the Tratnik polynomials of 
$q$-Krawtchouk type.

We have considered bivariate polynomials on finite grids but it should be possible to 
consider infinite grids with for example Wilson rather than Racah polynomials (see \textit{e.g.}~\cite{Trat-cont} for the Tratnik case). 

Finally, we would like to undertake the classification of all bivariate polynomials that are solutions of a bispectral problem. 
This would correspond to a generalization of the Leonard theorem  \cite{Leo}, which 
in the univariate case, led to the definition
and the study of Leonard pairs \cite{Ter}. This result would play an important role in the classification of the bivariate $P$- and $Q$-polynomial association schemes which have just been defined in \cite{bi,bannai}. 
The extension of the concept of Leonard pair to the bivariate case has been initiated in \cite{IT, CZ} with the introduction of 
(factorized) $A_2$- or $B_2$- Leonard pairs.
We intend to pursue these questions in the future.

\paragraph{Acknowledgements:}
N.~Cramp\'e is partially supported by the international research project AAPT of the CNRS. L.~Frappat and E.~Ragoucy are partially supported by Universit\'e Savoie Mont Blanc and Conseil Savoie Mont Blanc grant APOINT. E.~Ragoucy is partially supported by the CRM-Simons Scholars-in-residence programm. J.~Gaboriaud is supported by JSPS KAKENHI Grant Numbers 22F21320 and 22KF0189. The research of L.~Vinet is supported by a Discovery Grant from the Natural Sciences and Engineering Research Council (NSERC) of Canada. M.~Zaimi holds an Alexander-Graham-Bell scholarship from the NSERC.

\paragraph{Data availability statement:}
The data that supports the findings of this study are available within the article. 

\appendix
\section{Proof of relation (\ref{eq:Le1}) \label{App:A}}
\paragraph{The case $\epsilon=-1$.}
First look at relation \eqref{eq:Le1} for $\epsilon=-1$.
Modify the expressions \eqref{def:Amp}, \eqref{def:Am0} and \eqref{def:Amm}
of $A^{-,+}_{j-1,a}$, $A^{-,0}_{j-1,a}$ and $A^{-,-}_{j-1,a}$ using 
the following relations which are proven by direct calculations:
\begin{align}
\begin{aligned}
&F(-a-c_{12}-1;c_1,c_2)\,A^-_{j-1}(c_3,c_0,c_4;N-a+1)=  D^-(a;c_1,c_2;N-j+1)\,F(j-1;c_4,c_0)\,,\label{eq:le3}\\[1ex]
&\left[F(a;c_1,c_2)+F(-a-c_{12}-1;c_1,c_2)\right]\,
A_{j-1}(c_3,c_0,c_4;N-a)\\
&\qquad
= \left[-S^-(a;c_1,c_2;N-j+1)-\lambda^+(a;c_{12};N-j)\right]
\,F(j-1;c_4,c_0)\,,\\[1ex]
&F(a;c_1,c_2)\,A^+_{j-1}(c_0,c_4;N-a-1) = B^-(a;c_1,c_2;N-j+1)\,F(j-1;c_4,c_0)\,.
\end{aligned}
\end{align}  
Then, the l.h.s.~of \eqref{eq:Le1} becomes
\begin{align}
 F(j-1;c_4,c_0)\Big(& p_i(a-1;N-j)D^-(a;c_1,c_2;N-j+1)+p_i(a+1;N-j)B^-(a;c_1,c_2;N-j+1)\nonumber\\
 &+p_i(a;N-j)\big(-S^-(a;c_1,c_2;N-j+1)-\lambda^+(a;c_{12};N-j)\big)\Big)\,.\label{eq:tem1}
\end{align}
We recognize in \eqref{eq:tem1} the sum of the r.h.s.~of the contiguity difference equation \eqref{eq:cont-diff-rac} and the l.h.s.~of the contiguity recurrence relation \eqref{eq:cont-rec-rac}. Using both as well as
\begin{align}
  &    F(j;c_4,c_0)\,\mu^-_{i}(c_{23};N-j)=-A_j(c_1,c_0,c_4;N-i)\,,\label{eq:FmA}
\end{align}
we prove relation \eqref{eq:Le1} for $\epsilon=-1$.  
\paragraph{The case $\epsilon=-1$.} The proof follows the same lines as
the case $\epsilon=1$.
\paragraph{The case $\epsilon=0$.} The l.h.s.~of \eqref{eq:Le1} reads
\begin{equation}\label{eq:3.10}
\begin{split}
 &p_i(x;N-j)\,A_{j,x}^{0,0}(c_3,c_0,c_4,c_1)
 -\left[F(j;c_0,c_4)+F(-j-c_{04}-1;c_0,c_4)\right]\\
&\qquad \times\, \left[p_i(x-1;N-j)\,D(x;c_1,c_2,c_3;N-j) 
 +p_i(x+1;N-j)\,B(x;c_1,c_2,c_3;N-j)\right]\,,
\end{split}
\end{equation}
where we first used relations \eqref{def:A0p}, \eqref{def:A0m} and then 
the equations  
\begin{align}
&\left[F(x;c_2,c_1)+F(-x-c_{12}-1;c_2,c_1)\right]\,B(y;c_3,c_0,c_4;N-x)
= \Sigma^+_x(c_4,c_2,c_1;N-y-1) \,F(y;c_3,c_0)\,,\nonumber\\[1ex]
&\left[F(x;c_2,c_1)+F(-x-c_{12}-1;c_2,c_1)\right]\,D(y;c_3,c_0,c_4;N-x)\nonumber\\
&\qquad\qquad\qquad\qquad= \Sigma^-_x(c_4,c_2,c_1;N-y+1) \,F(-y-c_{03}-1;c_3,c_0)\,.
\end{align}    
Using difference equation \eqref{eq:diff-rac},  expression \eqref{eq:3.10} is rewritten as follows
\begin{align}\label{eq:3.11}
\begin{aligned}
&p_i(x;N-j)\,\Big(A_{j,x}^{0,0}(c_3,c_0,c_4,c_1)-\left[F(j;c_0,c_4)+F(-j-c_{04}-1;c_0,c_4)\right] \\
&\hspace{6cm}\times
\left[\mu_i(c_{23})+S(x;c_1,c_2,c_3;N-j)\right]\Big) \\
&=-p_i(x;N-j)\,\left[F(j;c_0,c_4)+F(-j-c_{04}-1;c_0,c_4)\right] \\
&\qquad\qquad\times\Big( \lambda(x;c_{12}) + i(i+c_{23}+1)+\tfrac{1}{2}(c_2+1)(c_{123}+1)\Big)\,.
\end{aligned}
\end{align}
The last line is obtained by a straightforward computation.
Finally, with the help of the recurrence relation \eqref{eq:rec-rac}, one gets the r.h.s.~of \eqref{eq:Le1}.
This ends the proof.

\end{document}